\documentclass[a4,amssymb,amsmath,aps,prb,twocolumn,superscriptaddress,showpacs,floatfix]{revtex4-1}
\usepackage{amssymb,amsmath}
\usepackage{color}
\usepackage{xcolor}
\usepackage{graphicx}
\usepackage{epstopdf}
\usepackage{bm}
\usepackage[space]{grffile}
\usepackage{hyperref}
\hypersetup{colorlinks=true, citecolor=blue, urlcolor=blue, linkcolor=red}

\newcommand{\lasriro}{(Sr$_{1-x}$La$_x$)$_3$Ir$_2$O$_7$}

\begin{document}
\title{Putative magnetic quantum criticality in (Sr$_{1-x}$La$_x$)$_3$Ir$_2$O$_7$}

\author{J. G. Vale}
\email{j.vale@ucl.ac.uk}
\affiliation{London Centre for Nanotechnology and Department of Physics and Astronomy, University College London, Gower Street, London, WC1E 6BT, United Kingdom}

\author{E. C. Hunter}
\affiliation{School of Physics and Astronomy, The University of Edinburgh,
James Clerk Maxwell Building, Mayfield Road, Edinburgh EH9 2TT, United Kingdom}
\affiliation{Inorganic Chemistry Laboratory, South Parks Road, Oxford, OX1 3QR, United Kingdom}
\begin{abstract}
(Sr$_{1-x}$La$_x$)$_3$Ir$_2$O$_7$ undergoes a bulk insulator-to-metal transition (IMT) at $x\approx0.04$. 
Through careful analysis of previously published data ($x=0.053,0.061,0.076$), 
we find an extended region below the Debye temperature in which the resistivity appears to scale linearly with temperature.
Meanwhile resonant (in)elastic x-ray scattering data ($x=0.065$) suggest a possible crossover from quantum paramagnetic to quantum critical phenomenology between 100 and 200~K. 
We put this into context with other results, and propose a possible phase diagram as a function of doping.
\end{abstract}
\maketitle

\begin{figure}
\includegraphics{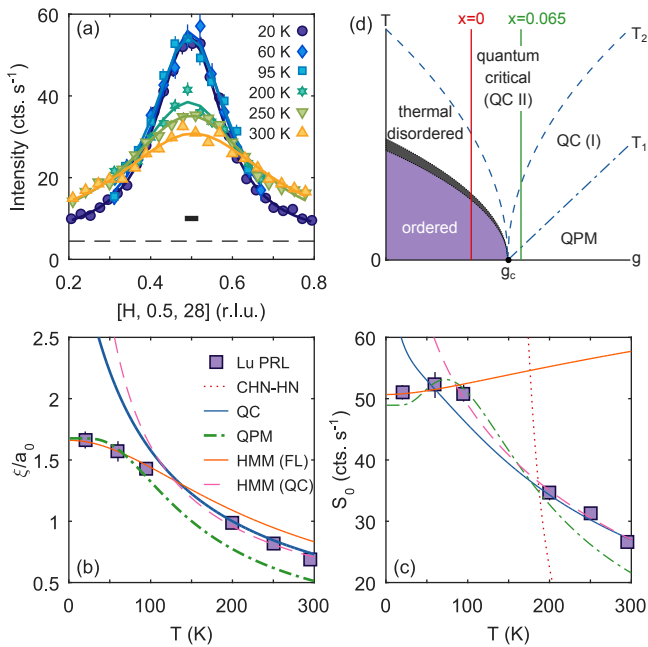}
\caption{Analysis of resonant x-ray magnetic scattering (REXS) data collected by Lu \emph{et al} for $x=0.065$.\cite{lu2017} (a): Intensity of $(\tfrac{1}{2},\,\tfrac{1}{2},\,28)$ magnetic Bragg peak as a function of temperature. Solid line is best fit to data, dashed line indicates linear background. Solid bar is FWHM of nearby structural Bragg peak. (b): Comparison of in-plane correlation length $\xi$ divided by Ir-Ir nearest-neighbour distance $a_0=\text{3.9~\AA}$ with various theoretical models. (c): Peak amplitude $S_0$. (d): Possible magnetic phase diagram for \lasriro\ as a function of doping and temperature. CHN-HN: Renormalized classical model.\cite{hasenfratz1991} QC: quantum critical.\cite{chubukov1994} QPM: quantum paramagnet (quantum disordered).\cite{chubukov1994} Orange solid (pink dashed): Hertz-Millis-Moriya model in Fermi liquid (quantum critical) regime.\cite{vonloehneysen2007} The CHN-HN, QC, and QPM models plotted in (b) and (c) use $\tilde{J}=\text{62~meV}$, with $\Delta=\text{14~meV}$ also used for the  QC and QPM models.}
\label{REXS_xi_S0}
\end{figure}

The properties of a system proximate to a quantum critical point (QCP) at $g=g_c$ generally fit the following scenario.
At low temperatures, and for $g<g_c$, the ground state may exhibit some form of long-range order (LRO) above its lower critical dimension. In antiferromagnets for instance, this corresponds to N\'{e}el order. This order is destroyed by classical thermal fluctuations, which dictate the scaling of thermodynamic properties in the vicinity of some critical temperature $T_c$. Above the transition, quasiparticles may still be well-defined on intermediate length scales, even though LRO has disappeared. This corresponds to a so-called thermal disordered regime.
At sufficiently high temperatures ($T\sim|g-g_c|^{\nu z}$), these quasiparticles are replaced by a critical continuum of excitations. This continuum is thermally excited; which leads to a characteristic $\omega/T$ scaling of the spin fluctuations in the vicinity of the critical wavevector, and unconventional power-law temperature dependences of physical observables.
If instead $g>g_c$, then the ground state is disordered and characterized by well-defined quasiparticle excitations. The properties of the system are dictated primarily by the magnitude of a singlet-triplet gap $\Delta\sim (g-g_c)^{\nu z}$, which exists at all wavevectors. We refer to this as a quantum paramagnetic state, although it is also known as quantum disordered behavior in the literature. A crossover to quantum critical behavior typically occurs around $T\sim \Delta$ (Fig.~\ref{REXS_xi_S0}d). 
Detailed reviews of quantum phase transitions (QPTs) are given in Refs.~\onlinecite{sachdev2011_QPT} and \onlinecite{vonloehneysen2007}, among others. 

A number of the cuprates have been proposed -- albeit with some controversy -- to undergo QPTs as a function of doping. These include the (hole-doped) high-temperature superconductor La$_{2-x}$Sr$_x$CuO$_4$ (LSCO), in which a magnetic QCP may lie underneath the superconducting dome.\cite{cooper2009, taillefer2010, sachdev2010}
Some similarities can be drawn between LSCO and the electron-doped perovskite iridate (Sr$_{1-x}$La$_x$)$_2$IrO$_4$. This material, like LSCO, is an insulator and easy-plane antiferromagnet below the N\'{e}el temperature at low carrier doping. It undergoes an insulator-to-metal transition (IMT) at $x\approx 0.04$,\cite{ge2011,chen2015}  with evidence of a pseudogap and hole-like Fermi surface in the metallic phase $x\geq 0.05$,\cite{kim2014_arpes, delatorre2015_sr214} along with possible spin density wave (SDW) order.\cite{chen2018} In contrast with LSCO, however, nanoscale electronic phase separation can be observed well into the metallic regime.\cite{chen2015} Moreover, experimental evidence of a QPT in this system is still outstanding.

Meanwhile the bilayer compound \lasriro\ has no direct analogues with any of the cuprates. Resistivity measurements have shown that \lasriro\ undergoes an IMT at $x=0.04$, \cite{li2013, hogan2015, hunterphd} similar to the single-layer material. Neutron scattering and second harmonic generation (SHG) measurements determined that the IMT is first order, with a structural phase transition occurring in the metallic phase at $T_s \approx \text{200~K}$.\cite{hogan2015,chu2017} Yet no pseudogap could be observed by angle-resolved photoelectron spectroscopy (ARPES), with small electron-like Fermi pockets present in the metallic phase.\cite{delatorre2014_sr327, he2015, affeldt2017} Furthermore, electronic phase separation does occur in the vicinity of the IMT, but disappears for $x>0.04$.\cite{hogan2015}
Magnetization and neutron scattering measurements show that N\'{e}el LRO disappears above the IMT.\cite{hogan2015} Resonant elastic x-ray scattering (REXS) data demonstrates, however, that short-ranged in-plane magnetic order persists up to 300~K,\cite{lu2017} at least for $x=0.065$.\footnote{Hogan \emph{et al.}~could not observe magnetic order on a sample with $x=0.05$ by REXS. Note that their measurements were performed on a dedicated resonant diffraction instrument, with a momentum resolution at least an order of magnitude better than the RIXS spectrometer used in Ref.~\onlinecite{lu2017}. This makes it more difficult to observe weak diffuse scattering.} 
The lack of dependence upon $L$ implies the loss of interlayer correlations deep in the metallic phase.
Resonant inelastic x-ray scattering (RIXS) measurements on the undoped compound reveal strongly gapped spin wave excitations; with an additional longitudinal mode interpreted as evidence of possible quantum dimer character and a proximate QPT.\cite{kim2012_sr327,moretti2015} Upon electron doping, the spin excitations become progressively more damped.\cite{hogan2016, lu2017} Whilst there are discrepancies between the two studies,\footnote{These are likely due to differences in calculation of the doping level $x$. One of the authors has noted that the value determined for $x$ can vary significantly from the nominal doping level;\cite{hunterphd} depending on the technique used, whether the sample was cleaved beforehand, and so on.} Lu \emph{et al.}~find that the spin gap collapses dramatically for $x=0.065$, proposing that the 2D behaviour also extends to the dynamics.

We examine some of the previously published experimental data for \lasriro\ in more detail. What we find are three distinct electronic regimes for $x>0.05$, which can be discriminated through clear gradient changes in the resistivity as a function of temperature. These correlate well with phase boundaries determined by other techniques. Meanwhile REXS data suggests a possible crossover between quantum paramagnetic and quantum critical behavior between 100 and 200~K. This is corroborated by apparent $\omega/T^{\alpha}$ scaling observed in RIXS data. Each of these shall be discussed in turn, starting with the temperature dependence of the resistivity.
\begin{figure}
\includegraphics{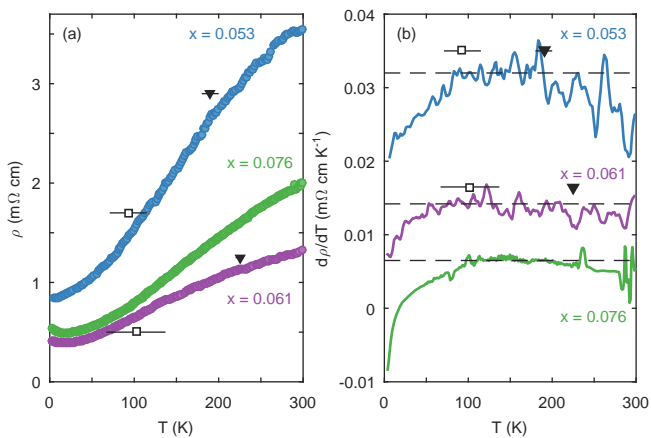}
\caption{(a): Summary of resistivity data from Refs.~\onlinecite{hogan2015} ($x=\text{0.053, 0.061}$) and \onlinecite{hunterphd} ($x=\text{0.076}$). 
Filled triangles indicate the structural phase transition at $T_s$.\cite{chu2017} Open circles: loss of coherent quasiparticle spectral weight at $T_{\mathrm{coh}}$ determined from ARPES.\cite{affeldt2017}
(b): Derivative $\mathrm{d}\rho/\mathrm{d}T$ of the same data; smoothed with an unweighted 5 point moving average and offset by a constant scale factor for clarity. Dotted line: guide to the eye highlighting region where $\mathrm{d}\rho/\mathrm{d}T$ is approximately constant. All other annotations are the same as displayed in (a).}
\label{resistivity}
\end{figure}

In Fig.~\ref{resistivity}(a), the resistivity of three different samples with $x=\text{0.053(10), 0.061(10), and 0.076(11)}$, has been plotted as a function of temperature. There are three immediately apparent observations from the data. The first is that all three samples are clearly metallic up to 300~K. 
The low temperature upturn present for the $x=\text{0.061}$ and $x=\text{0.076}$ samples can be attributed to Anderson localization, likely due to sample inhomogeneity. Moreover, the $x=\text{0.061}$ sample has a consistently lower resistivity that the other two samples. It has been noted, however, that sample to sample variation within a batch can lead to differences in the absolute magnitude of the resistivity by a factor of 2.\cite{hunterphd} The corresponding residual resistivity ratio (RRR) is about 4 for all three samples, which is a direct consequence of the flux growth method used to generate single crystals of this material. In this sense, \lasriro\ is a somewhat dirty system.

A less obvious observation is that a kink is evident at ca.~200~K, which we note is a similar temperature to the structural phase transition observed via neutron scattering in Ref.~\onlinecite{hogan2015}. This is more clearly displayed in Fig.~\ref{resistivity}(b), in which we plot the temperature derivative of the resistivity, $\mathrm{d}\rho/\mathrm{d}T$. A further change in slope occurs around 100~K, which coincides remarkably well with the loss of coherent quasiparticle spectral weight at $T_{\mathrm{coh}}$ observed via ARPES.\cite{affeldt2017} 
Between these two temperatures, $\mathrm{d}\rho/\mathrm{d}T$ is approximately constant, implying $\rho \propto T$. 
Such behavior is expected for good metals well above the Debye temperature $\Theta_D$ due to electron-phonon scattering. Yet specific heat measurements reveal that the Debye temperature $\Theta_D \approx \text{270~K}$, and is essentially independent of doping.
We also note that the absolute value of the resistivity at 200~K is comparable to the Mott-Ioffe-Regel (MIR) limit, within which the mean free path $l$ is on the order of the lattice constant $a$.\cite{deng2013, he2015} This implies that the system is close to the so-called bad metal regime. Hence we can rule out phonons as the leading cause for $T$-linear resistivity in \lasriro\, and our observations are likely indicative of alternative phenomenology. 
One possibility is that the linear scaling of the resistivity above $T_{\mathrm{coh}}$ may be representative of underlying quantum critical behavior. In the cuprates for instance, the $\rho \propto T$ scaling present in the ``strange metal'' phase has been suggested to manifest due to scattering from some fluctuating order parameter. 

If \lasriro\ does indeed undergo a QPT, then signatures of this should also be seen in the magnetic behavior.
As mentioned previously, REXS measurements by Lu \emph{et al.} \cite{lu2017} reveal short-ranged magnetic order which persists above the IMT for $x>0.04$. 
The magnetic $(0.5,0.5,28)$ Bragg peak appears to weaken and broaden with increasing temperature, however significant in-plane correlations are still observable at 300~K. Hogan \emph{et al.~}also found with RIXS that the magnon peak was invariant with $L$ for their $x=0.07$ sample (within experimental uncertainty).\cite{hogan2016}
Noting that two-dimensional layered materials frequently exhibit such behavior, we proceeded to examine the data in more detail [Fig.~\ref{REXS_xi_S0}(a)].

We fitted the data at each temperature to a Voigt function, with the Gaussian component fixed to the width of a typical structural Bragg reflection, in order to approximate the instrumental resolution function. Varying this width within sensible bounds does not change our results significantly. Additionally, the background was fixed at all temperatures to that obtained from fitting at 20~K. 
What can be seen is that the correlation length $\xi$ and equal-time structure factor $S_0$ are approximately constant up to 100~K [Fig.~\ref{REXS_xi_S0}(b,c)], with these parameters decreasing at higher temperatures. There thus appear to be two distinct temperature regimes within the data.

Our findings shall initially be discussed in terms of the $\mathcal{O}(N)$ quantum non-linear sigma model (QNL$\sigma$M); probably the simplest model to undergo a continuous quantum phase transition (QPT) in $2+1$ dimensions. Specifically, we use key results given within Ref.~\onlinecite{chubukov1994}, which have been obtained through exact solution in the $N=\infty$ limit.
Corrections to order $1/N$ are non-trivial to calculate for $g>g_c$, which is why they have been neglected in this initial study. We note at this point that further-neighbour interactions (and anisotropies) are important for \lasriro, which clearly manifest in the observed spin wave dispersion.\cite{kim2012_sr327, moretti2015, hogan2016, lu2017} Therefore, unless otherwise stated, the following discussion applies to an effective exchange interaction $\tilde{J}$ which includes the effects of the other terms in the Hamiltonian.

The QNL$\sigma$M exhibits LRO at $T=0$, provided that the coupling $g = \hbar c\sqrt{2\pi}/(k_B\rho_sa)<g_c$, where $c=2\sqrt{2}Z_cS\tilde{J}a$ is the spin wave velocity, $a$ is the lattice constant, $\rho_s$ is the spin stiffness, and $Z_c=1+\eta$ is a renormalization factor which describes the effect of quantum fluctuations. From now on we take $Z_c=1$, in order to better compare with the experimental results.  Real materials typically order at non-zero temperatures as a consequence of weak anisotropies or further-neighbour interactions. Given the intrinsically broad magnetic Bragg peak at 30~K, and that bulk susceptibility data shows paramagnetic behavior at all temperatures, then this implies that the material lies on the $g>g_c$ side of the putative QPT for $x=0.065$.

In this regime, both $\xi$ and $S_0$ are expected to scale with $x_2=k_BT/\Delta$, where $\Delta$ is the singlet-triplet gap:
\begin{align}
\xi^{-1} &= \frac{k_BT}{\hbar c}X_2(x_2)	\nonumber \\
S(k\rightarrow 0) &\propto \frac{(\hbar c)^2}{k_BT}\frac{x_2 \coth(X_2/2)}{2X_2(x_2)}.
\end{align} 
The parameter $X_2$ in the preceding expressions is a scaling function, which has the following asymptotic limits in the quantum paramagnetic ($x_2\ll 1$) and quantum critical ($x_2\gg 1$) regions:
\begin{align}
X_2(x_2) &= x_2^{-1} + 2e^{-1/x_2}, \quad x_2\ll 1 \nonumber \\
         &= 2\ln{\left(\frac{\sqrt{5}+1}{2}\right)}+\frac{1}{\sqrt{5}x_2}, \quad x_2\gg 1
\end{align}
Hence this implies a correlation length which is on the order of $\Delta^{-1}$ (for $x_2\ll 1$) or $T^{-1}$ (for $x_2\gg 1$).

This theoretical model is compared with the experimental data in Figs.~\ref{REXS_xi_S0}(b) and \ref{REXS_xi_S0}(c).
We find that the quantum paramagnetic model describes the experimental data below 100~K quite well. At higher temperatures however, it appears to underestimate the correlation length and structure factor. Meanwhile the quantum critical model agrees with the data above 200~K, but diverges at lower temperatures.  This suggests that there may be a crossover between the two regimes in the temperature range 100--200 K. Unfortunately, there is currently no experimental data available which corresponds to this region.

The agreement between the experimental data and theory is also quantitative. Note that the value of $\Delta$ we obtain ($\Delta=\text{14~meV}$) is comparable with the magnon gap observed in RIXS for $x=0.065$.\cite{lu2017}
Furthermore, $\tilde{J}=\text{62~meV}$ is in excellent agreement with the effective nearest-neighbor coupling derived from the RIXS data: $\tilde{J}=\sum_i J_i z_i=\text{64~meV}$, where $J_i$ are the individual coupling parameters (including anisotropies), and $z_i$ the number of neighbors.
Whilst there are some question marks regarding the quantitative mapping of the QNL$\sigma$M to $S=1/2$ Heisenberg spin systems (for example, see the discussion in Ref.~\onlinecite{dai2005}), the correlation is, nevertheless, remarkable.

We also plot the expected temperature dependence of the correlation length and structure factor within the Hertz-Millis-Moriya model (HMM) for 3D nearly antiferromagnetic metals ($d=3$, $z=2$). In the HMM picture, spin fluctuations become soft at the QCP, and are damped by a background of itinerant electrons. This is somewhat related to the paramagnon (SCR) theory by Moriya.\cite{moriya1985}
At low temperature ($T<T^*$), the inverse correlation length $\xi^{-1}$ exhibits the $T^2$ dependence characteristic of a Fermi liquid. Above $T^*$, the system exhibits quantum criticality. It has been determined that $\xi^{-2} = |g-g_c| + AT^{2/3}$, with the first term dominating in region I, and the second in region II (Fig.~\ref{REXS_xi_S0}d).
Meanwhile the equal-time structure factor $S_0$ is given by:
\begin{equation}
S_0(k\rightarrow 0) = \frac{1}{\pi}\int^{\infty}_{-\infty} \! [n(\omega)+1]\,\mathsf{Im}\left[\frac{A\xi^2}{1 - i\omega/\omega_{SF}}\right]\hbar\omega,
\end{equation}
where $n(\omega)=(e^{\hbar\omega/k_BT}-1)^{-1}$, and $\omega_{SF}$ is the characteristic energy for spin fluctuations. We assume that $\omega_{SF}=\text{20~meV}$, which corresponds to the experimental spin wave energy at 30~K, and is fixed as a function of temperature.
Again, we observe that the data below 100~K are well described by the low temperature predictions of the HMM model, with the higher temperature data more representative of quantum critical phenomenology (mostly region II). 

There is further evidence that the magnetic fluctuations at high temperature may be indicative of quantum critical behavior. In the vicinity of a QCP, spin fluctuations at the antiferromagnetic (AFM) wavevector $\bm{Q}_{\mathrm{AF}}$ are expected to exhibit $\omega/T^\alpha$ scaling, where $\alpha$ is an independent scaling exponent. Different models for the criticality predict different results. For instance, in the HMM model, spin fluctuations are dominated by the AFM order parameter. 
Consequently for a 3D nearly AFM metal, it predicts $E/T^{3/2}$ scaling of the dynamic spin susceptibility: $\chi^{-1}(\bm{Q}_{\mathrm{AF}},E,T) = a^{-1}(T^{3/2}-ibE)$. The parameter $b$ is related to the characteristic energy of spin fluctuations $\omega_{\mathrm{SF}}$ defined earlier.
Meanwhile anomalous exponents have been observed in some heavy fermion systems, which are believed to correspond to ``local criticality''. Schr\"{o}der \cite{schroeder2000} and Poudel \cite{poudel2017} have proposed a modified Curie-Weiss law to describe the inverse dynamic susceptibility in such systems: $\chi^{-1}(\bm{q},E, T) = c^{-1}[\theta^{\alpha}+(T-iE)^\alpha]$, where $\theta(\bm{Q}-\bm{Q}_{\mathrm{AF}})$ captures the wave-vector dependence of the magnetic fluctuations similar to the Curie-Weiss temperature. Note that in the latter picture, the fluctuations become critical in the time domain everywhere in $\bm{q}$, rather than just at $\bm{Q}_{\mathrm{AF}}$.

\begin{figure}[t]
\includegraphics{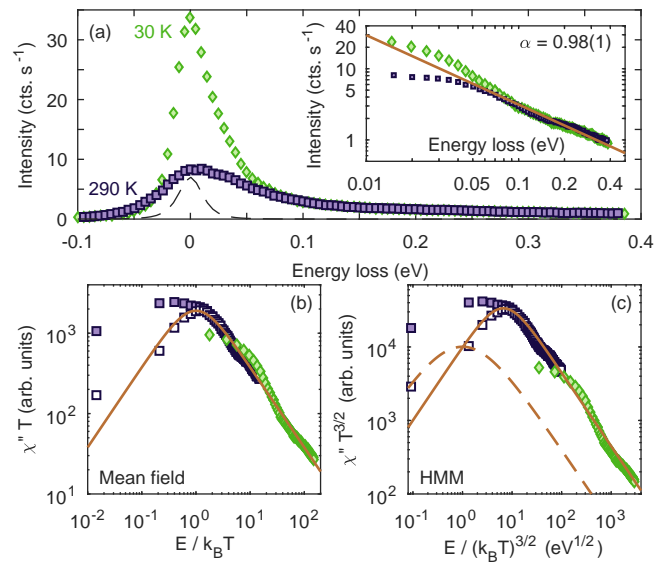}
\caption{Temperature dependence of RIXS spectra at (0.5, 0.5, 26.5) for $x=0.065$. (a): Comparison of data at 30~K (green diamonds) and 290~K (purple squares). Dashed line: elastic line at 290~K.
Reproduced from Fig.~S3(a) of Ref.~\onlinecite{lu2017}. Inset shows same data plotted on double logarithmic axes, along with best fit to power law with exponent $\alpha=0.98(1)$.
(b,c): Scaling plots of the imaginary part of the dynamic susceptibility $\chi''(E,T)$, plotted as $\chi''T^{\alpha}=f(E/T^{\beta})$, with (b) $\alpha=\beta=1$ or (c) $\alpha=\beta=3/2$. Note $\chi''$ has been obtained from the data presented in (a) using the relation $\chi''(E,T)=I[1-\exp{(E/k_BT)}]$. The scaling functions $f(E/T^{\beta})$ are defined in the main text. Open symbols reflect 290~K data with the elastic line subtracted off. Solid lines are fits to the data. Dashed line in (c) uses $b=1$.}
\label{RIXS}
\end{figure}

In Fig.~\ref{RIXS}(a), we plot resonant inelastic x-ray scattering (RIXS) data at $(\pi,\pi)$ for $x=0.065$, which was previously published in Ref.~\onlinecite{lu2017}. The spin excitations appear highly damped, and extend to 0.4~eV energy loss at both 30~K and 295~K. Moreover, when we plot the data on double logarithmic axes [inset of Fig.~\ref{RIXS}(a)], it appears to scale approximately linearly for $E>\text{0.05~eV}$. This implies that the dynamic critical exponent $z=2/\alpha \sim 2$. Such a value is expected both for three-dimensional nearly AFM metals,\cite{vonloehneysen2007} and quantum dimer models on a square lattice.\cite{isakov2011} One complication is that the ideal QNL$\sigma$M assumes Lorentz invariance, and hence $z=1$. However the presence of disorder (caused by doping) can break this invariance and give rise to $z\neq 1$.\cite{chubukov1994}

Scaling plots in Fig.~\ref{RIXS}(b) and \ref{RIXS}(c) compare the experimental data to the theoretical predictions for the mean field ($\alpha=1$) and HMM ($\alpha=3/2$) models respectively. Broadly speaking, both models provide an adequate description of the data. Some discrepancies at low $E/T$ can be overcome by subtracting off the resolution-limited elastic line and low-energy phonon contributions, the latter being observed in the undoped compound at 25~meV (compare filled and open symbols). 
We find that the best fit for the HMM model is obtained with $b=0.14$, significantly smaller from the expected value of unity. Note that the simple model given here is defined precisely at the QCP and AFM wavevector, and assumes zero anisotropy. Yet the REXS data presented earlier suggests that the putative magnetic QCP lies at $x<0.065$. Furthermore, the finite momentum resolution of the RIXS spectrometer means that we sample a number of momentum transfers close to $(\pi,\pi)$. Finally, the 20~meV spin gap present at 30~K may persist to some degree at high temperatures. Such a deviation, is therefore, not entirely unexpected.

\begin{figure}
\includegraphics{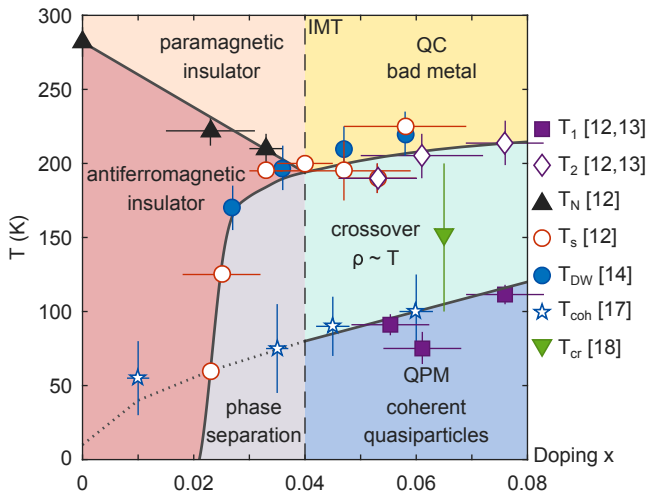}
\caption{Possible unified phase diagram for \lasriro\ summarising transition temperatures observed with various techniques. Black triangles (red open circles): N\'{e}el temperature $T_{\mathrm{N}}$ (structural transition temperature $T_{\mathrm{s}}$) from neutron powder diffraction.\cite{hogan2015} Blue filled circles: onset of putative CDW $T_{\mathrm{DW}}$ from pump-probe optical reflectivity. \cite{chu2017} Green filled triangles: crossover temperature from REXS. \cite{lu2017} Blue stars: loss of coherent spectral weight in ARPES.\cite{affeldt2017} Filled squares (open diamonds): Onset (end) of constant $\mathrm{d}\rho/\mathrm{d}T$ at $T_1$ ($T_2$).\cite{hogan2015, hunterphd}}
\label{phase_diagram}
\end{figure}
We conclude by putting the analysis presented here in context with other experimental results. Specifically, we extend the temperature-doping phase diagram to include recent ARPES data, and our resistivity and REXS results (Fig.~\ref{phase_diagram}). 
A striking correlation is apparent between the onset of constant $\mathrm{d}\rho/\mathrm{d}T$ at $T_1$, and the loss of coherent spectral weight at $T_{\text{coh}}$ observed in ARPES.\cite{affeldt2017} A further change in slope of the resistivity at $T_2$ also appears to coincide with the structural phase transition observed by neutron scattering,\cite{hogan2015} and the loss of putative charge density wave (CDW) ordering from pump-probe optical reflectivity data.\cite{chu2017} Whether the $\rho\propto T$ behaviour persists above $T_s$ is still uncertain within the limits of the data presented here.
We note the apparent similarity in the transport behavior as observed for overdoped cuprates.\cite{hussey2008} In the cuprates, however, the quasiparticle peak (from ARPES) persists into the $\rho\propto T$ regime.

The REXS results also show a possible crossover between quantum paramagnetic and quantum critical phenomenology somewhere between 100 and 200~K. This is consistent with our value of $T^*$, which implies that the Hertz-Millis-Moriya picture (Fig.~\ref{REXS_xi_S0}d) may be relevant for \lasriro. Yet the (limited) RIXS data is less clear cut. There appears to be $E/T^{\alpha}$ scaling of the dynamic spin susceptibility, as would be expected in the vicinity of a QCP. At present, it is not possible to conclusively distinguish between the HMM or simple mean-field pictures. Even so, it demonstrates that quantum criticality seems to extend to the spin dynamics.

Clearly there is a significant difference in the electronic and magnetic behavior of \lasriro\ compared to its single layer counterpart (Sr$_{1-x}$La$_x$)$_2$IrO$_4$. What the data presented here show is that structural, electronic, and magnetic degrees of freedom are directly coupled in \lasriro, giving rise to a rich phase diagram, and potentially containing a hidden magnetic QCP. 

A number of outstanding questions remain. The first is the nature of the apparent crossover between the quantum paramagnetic and quantum critical regimes. This can only be definitively answered through collection of more data -- both in the elastic and inelastic channels -- between 100~K and 200~K at various doping levels. This in turn leads to whether the $E/T^{\alpha}$ scaling is universal as a function of temperature and wavevector. If so, then this would imply a departure from the HMM picture, which is only expected to be relevant in the vicinity of $\bm{Q}_{\mathrm{AF}}$. 
At present it remains unclear whether there is a single zero temperature QCP, or a line of first order IMTs which end at a finite temperature critical end point.\cite{vucicevic2013, vucicevic2015} Magnetic susceptibility measurements as a function of pressure may help to discriminate between these two scenarios.
It is also curious that the nominally three-dimensional \lasriro\ appears to map onto the 2+1 dimensional QNL$\sigma$M, yet the single-layer (Sr$_{1-x}$La$_x$)$_2$IrO$_4$ does not (for low doping at any rate). This may be related to the electronic phase separation prevalent in the single layer compound above the IMT. Nevertheless, it may be worth revisiting (Sr$_{1-x}$La$_x$)$_2$IrO$_4$ to look for evidence of quantum criticality in the metallic phase.

\acknowledgements{This work was supported by the EPSRC (grants EP/N027671/1, EP/N034872/1, EP/P013449/1). We thank X. Lu for supplying the REXS and RIXS data, and D.~F. McMorrow for a critical reading of the manuscript.}

%
\end{document}